\documentclass[a4paper, fleqn, usenatbib, useAMS]{mnras}

\def\NU{N_{\rm UDG}}
\def\Mc{M_{\rm host}}
\def\Mh{M_{\rm sub}}

\def\aap{AA}
\def\apjl{ApJL}
\def\apjs{ApJS}
\def\mnras{MNRAS}
\def\apj{ApJ}
\def\aj{AJ}

\def\araa{ARAA}
\usepackage{afterpage}
\usepackage{multicol}        
\usepackage{graphicx}
\usepackage{float}
\usepackage{bm} 
\usepackage{amssymb}
\usepackage{hyperref}
\usepackage{amsmath} 
\usepackage{subfig} 
\usepackage{pdflscape}
\usepackage[export]{adjustbox}
\usepackage{mathrsfs} 
\usepackage{mathtools}

\title[The halo masses of UDGs]{The virial mass distribution of ultra-diffuse galaxies in clusters and groups}
\author[N. C. Amorisco et al.]{N. C. Amorisco$^{1, 2}$\thanks{E-mail:
nicola.amorisco@cfa.harvard.edu}\\
$^{1}$Max Planck Institute for Astrophysics,  Karl-Schwarzschild-Strasse 1, 85748 Garching, Germany \\
$^{2}$Institute for Theory and Computation,  Harvard-Smithsonian centre for Astrophysics,  60 Garden St.,  Cambridge,  MA 02138,  USA}

\begin{document}



\maketitle

\label{firstpage}

\begin{abstract}
We use the observed abundances of ultra-diffuse galaxies (UDGs) in clusters and groups and $\Lambda$CDM subhalo
mass functions to put constraints on the distribution {of present-day halo masses of satellite} UDGs. 
If all of the most massive subhaloes in the cluster host a UDG, UDGs occupy all subhaloes with 
$\log\Mh/M_\odot\gtrsim11$.
For a model in which the efficiency of UDG formation is higher around some characteristic halo mass, higher fractions 
of massive UDGs require larger spreads in the UDG mass distribution. {In a cluster with a virial mass of $10^{15}M_\odot$},
the 90\% upper limit for the fraction of 
UDGs with $\log\Mh/M_\odot>12$ is 7\%, occupying 70\% of all cluster subhaloes above the same mass.
To reproduce the observed abundances, however, the mass distribution of {satellite UDGs} has to be broad, with 
$>30\%$ having $\log\Mh/M_\odot<10.9$. 
This strongly supports that UDGs are part of a continuous distribution in which a majority are hosted by low mass haloes. 
The abundance of {satellite UDGs} may fall short of the linear relation with the cluster/group mass $\Mc$ in low-mass hosts, $\log \Mc/M_\odot\sim 12$. Characterising these deviations -- or the lack thereof -- will allow for stringent constraints on the UDG virial mass distribution.
\end{abstract}

\begin{keywords}
galaxies: dwarf --- galaxies: structure --- galaxies: formation --- galaxies: haloes  ---  galaxies: clusters \end{keywords}

\section{Introduction}

The suggestion that surface brightness limited surveys may significantly underestimate the 
total number of galaxies is at least twenty years old. Using a very simple model based on a 
standard $\Lambda$CDM framework, \citet{JD97} predicted the existence of an extremely 
abundant population of low surface brightness (LSB) galaxies, potentially extending to surface 
brightness levels of $\mu\gtrsim 30~$mag/arcsec$^2$. Reaching these depths remains exceptionally
challenging and great effort is currently being devoted to push to ever fainter limits, to probe a
yet largely unexplored regime of the galaxy formation process.

{LSB galaxies, including faint cluster galaxies, have been the object of numerous studies 
\citep[e.g.,][]{BB85,HF89,JD94,IB97,CC02,CC03,CA06,SP09,LF12,HY12}}, and the significant abundance of Ultra-diffuse galaxies (UDGs) in clusters \citep[e.g., ][]{vD15,Koda15,Mun15,vdB16,RT17a,Ja17,Le17,Ve17} 
has confirmed the above prediction. With mean surface brightnesses within their effective radius $R_e$ of
$\langle\mu\rangle\gtrsim 24.5~$mag/arcsec$^2$, UDGs have stellar masses of dwarf galaxies ($\log M_*/M_\odot\sim {7.5\div8.5}$), 
but $R_e >1.5~$kpc, quite larger than what is common among bright galaxies with similar stellar mass.
Within the framework considered by \citet{JD97}, UDGs owe their remarkable sizes to the high 
angular momentum of their dark matter halos \citep[see also e.g.,][]{MMW98,AD07}. Although simplistic, 
this scenario reproduces the abundance of UDGs in clusters and their size distribution \citep{NA16a},
and predicts that most UDGs are hosted by low-mass halos \citep[$\log \Mh/M_\odot\sim {10.3\div11.3}$ for a normal 
stellar-to-halo mass relation,][]{NA16a}. 
Stellar feedback has been proposed as alternative cause of the UDGs extended sizes \citep{AD16},
which would also require that most UDGs are hosted by low-mass haloes ($\log \Mh/M_\odot\sim11$).
Galaxies with the properties of UDGs have been obtained in recent hydrodynamical simulations \citep{AD16,Ch17},
but clear predictions for their population properties are not yet available. 

It remains possible, however, that a fraction of UDGs is hosted 
by haloes that are considerably more massive than suggested by the scenarios above, and 
potentially as massive as the Milky Way (MW) halo. 
{It has been proposed that some UDGs may be `failed' $L_*$ galaxies, with a different formation pathway.
Failure could be caused by gas stripping and/or extreme feedback processes, that may have halted their star formation.
UDGs would then fall significantly short of the stellar-to-halo mass relation and be hosted by `over-massive haloes' \citep[e.g.,][]{vD15,vD16,MB16a}.}
The only direct measurement of the virial mass of UDGs, based on the stacked weak lensing signal of $>700$ systems, 
can not rule out this possibility \citep{Sif17}. 
Indirect arguments on the mass of UDG hosting haloes appear to confirm that a majority have low mass 
haloes \citep[$\lesssim 2\times 10^{11}~M_\odot$][]{MB16a,MB16b,Pe16,NA16b,RT17a}. A few notable 
exceptions could however be interpreted as due to a fraction of systems with higher virial masses.
These exceptions include the high GC abundances of some Coma UDGs \citep[][together with some less extended 
LSB galaxies, Amorisco et al. 2016]{vD16,vD17}
and the central stellar velocity dispersion of a couple of Coma UDGs \citep{vD16,vD17}. In fact, 
if UDGs comply with the same scaling relations of normal galaxies, the heterogeneity of   
surface brightness values and sizes would suggest a mix of halo masses \citep{Za16b}.

It is therefore important to try and constrain the {\it distribution} of UDG virial masses. In this Letter, {we concentrate
on satellite UDGs, and constrain their present-day subhalo masses using literature measurements of their cluster and group abundances} \citep{Koda15,Mun15,vdB16,vdB17,RT17a,RT17b,Ja17}. 
Section~2 describes the data 
and our {simple} model. Section~3 details our statistical analysis and collects results. 
Section~4 discusses them and lays out the Conclusions.

\begin{figure}
\centering
\includegraphics[width=.95\columnwidth]{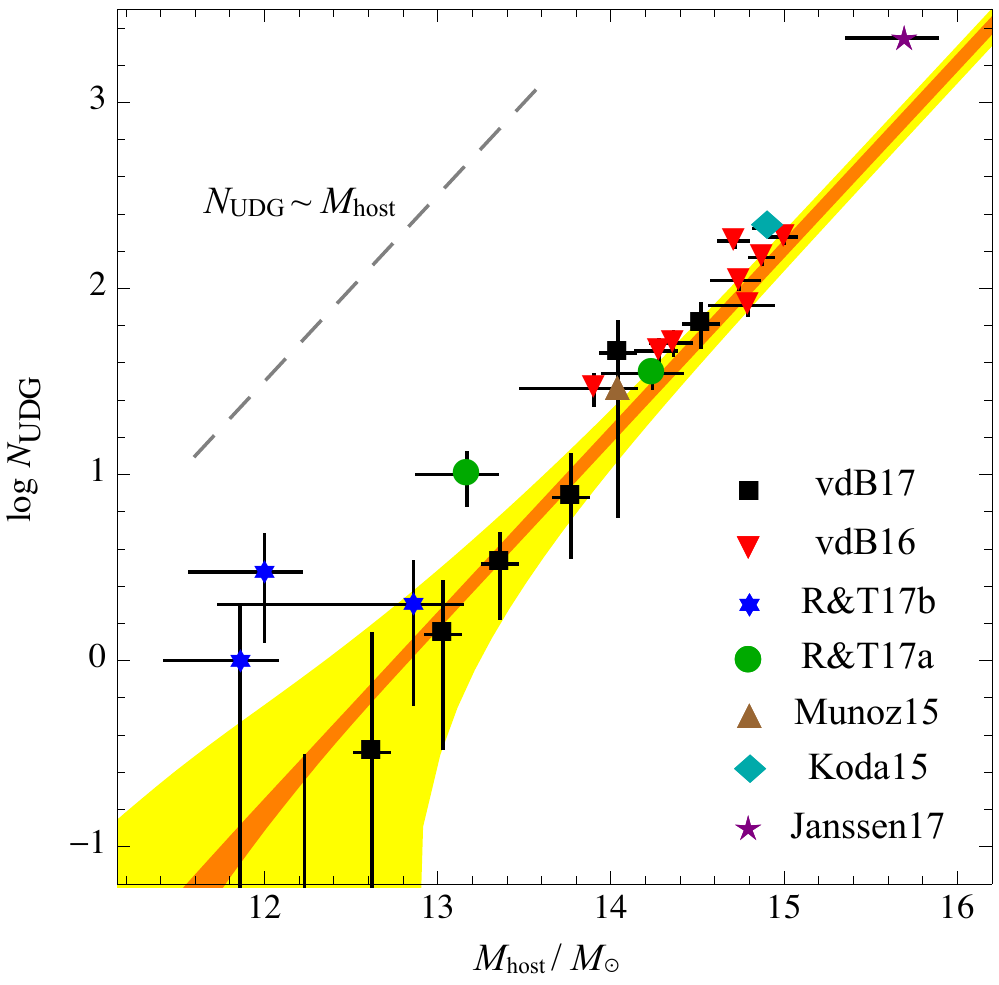}
\caption{UDG abundances in groups and clusters as measured by \citet{Koda15,Mun15,vdB16,vdB17,RT17a,RT17b,Ja17}.
 The dashed line shows the slope of a linear relation. The orange shading displays the 10-to-90\% confidence 
 region for the mean $\NU$-$\Mc$ relation obtained from our analysis, the yellow shaded region shows the 1-$\sigma$ scatter around it.}
\label{model}
\end{figure}

\section{The abundance of UDGs}

Significant effort has been put into measuring the abundance
of UDGs in galaxy clusters \citep[e.g., ][]{vdB16,RT17a,Ja17,Le17,Ve17}. These results have confirmed the initial
finding of \citet{vdB16} that the relation between the number of UDGs, $\NU$, and the virial mass of the host cluster, $\Mc$,
is compatible with being linear. 
{As discussed in \citet{NA16b}, an approximately linear relation between $\NU$ and $\Mc$ is straightforwardly 
reproduced if the following two conditions are satisfied: i) the physical mechanism that is responsible for the 
properties of UDGs is independent of environment,
ii) the majority of satellite UDGs have low-mass haloes. 
For small values of the subhalo-to-host mass ratio, $\Mh/\Mc\ll1$, the mean subhalo abundance per unit parent mass is independent of the mass of the host 
\citep[e.g.,][]{Gao04,vdB05,CG08}. Therefore, the two conditions above are sufficient
to ensure an approximately linear relation between $\NU$ and $\Mc$.}
  
Interestingly though, abundant UDG population have been detected in galaxy groups \citep[hereafter RT17 and vdB17]{RT17b,vdB17}, 
extending the same approximately linear relation valid in massive clusters \citep[see also][]{IT17}. Figure~1 reproduces the 
collection of data
presented by vdB17: over $\sim$4 orders of magnitude in $\Mc$, $\NU$ is approximately proportional to 
the mass of the host (the dashed line shows the slope of a linear relation in this plane). As we will show, these 
high abundances put strong constraints on the distribution halo masses of {satellite UDGs}, and so does 
the apparent linearity of this relation. The subhalo abundance per unit parent mass is not independent of 
parent mass when $\Mh/\Mc\sim1$: abundances are strongly suppressed for subhaloes above $\Mh/\Mc\sim0.1$ 
\citep[e.g.,][]{Gao04,CG08}. 
This causes potentially observable deviations from linearity in the relation between $\Mc$ and $\NU$ when the mix 
of UDG halo masses includes a significant fraction of massive haloes. We formalise these concepts in the 
following.

\subsection{Model UDG abundances}

We model the mean differential mass function of subhaloes with mass $\Mh$, $\langle N(\Mh) \rangle$, 
in a parent halo or cluster of mass $\Mc$ with a fitting function based on: i) the results of \citet{BK10} (hereafter BK10) 
and ii) the mentioned independence of the subhalo abundance per unit parent mass on the parent mass itself. 
The subhalo mass function in MW-mass haloes ($12\leq\log \Mc/M_\odot\leq12.5$) has been measured with high precision 
by BK10, using the Millennium-II Simulation \citep{BK09}.
We adopt the functional form and parameters suggested by these authors (their eqn.~(8)) and refer to this function as
\begin{equation}
\left. {{{\rm d}\langle N\rangle}\over{{\rm d}\log \Mh}}\right|_{\Mc=M_{\rm MW}}(\Mh)=\mathcal{N}(\Mh)\ .
\label{subhmfBK}
\end{equation}
This is a power-law with index\footnote{{To explore how any uncertainty on the slope $a$ influences our results
we also consider a mass function with $a=-1.86$ \citep[e.g.][]{Ji16,Bo16}. Other parameters are unchanged and we
impose that the number of subhaloes with mass $\Mh>10^{-4} M_{\rm MW}$ is the same.}}
  $a=-1.935$ for $\Mh/M_{\rm MW}\ll 1$ with an exponential truncation 
for subhalo masses $\Mh\gtrsim10^{11}M_\odot$.  
To calculate the differential mass function of subhaloes 
with mass $\Mh$ in any parent of mass $\Mc$ we scale eqn.~(\ref{subhmfBK}) using that the subhalo 
mass function is independent of the parent mass:
\begin{equation}
{{{\rm d}\langle N\rangle}\over{{\rm d}\log \Mh}}={\Mc \over {M_{\rm MW}}} \mathcal{N}\left(M_{\rm MW} {\Mh\over\Mc}\right){\mathcal{N}(m_0)\over{\mathcal{N}(M_{\rm MW}~ {m_0/\Mc})}}\ .
\label{subhmf}
\end{equation}
This uses that the shape of the truncation at high subhalo to parent mass ratio is also independent of the parent mass 
\citep[see e.g.][]{CG08}. 
In eqn~(\ref{subhmf}), $m_0$ is any subhalo mass satisfying $m_0\ll M_{\rm MW}$. {The mass function $\mathcal{N}$
has been measured using all central haloes with virial mass between $10^{12}M_\odot$ and $10^{12.5}M_\odot$. We 
therefore estimate the mean value $M_{\rm MW}$ (needed in eqn.~(\ref{subhmf})) and the relative uncertainty 
(necessary for our statistical analysis, see the following section) using a standard halo mass 
function with a slope of $-1.9$, which returns $\log M_{\rm MW}/M_\odot=12.23\pm0.14$.} 
For any cluster or group, eqn.~(\ref{subhmf}) allows us to calculate the mean number of subhaloes in any given mass 
interval. A fraction of these subhaloes will host UDGs.

\begin{figure}
\centering
\includegraphics[width=.95\columnwidth]{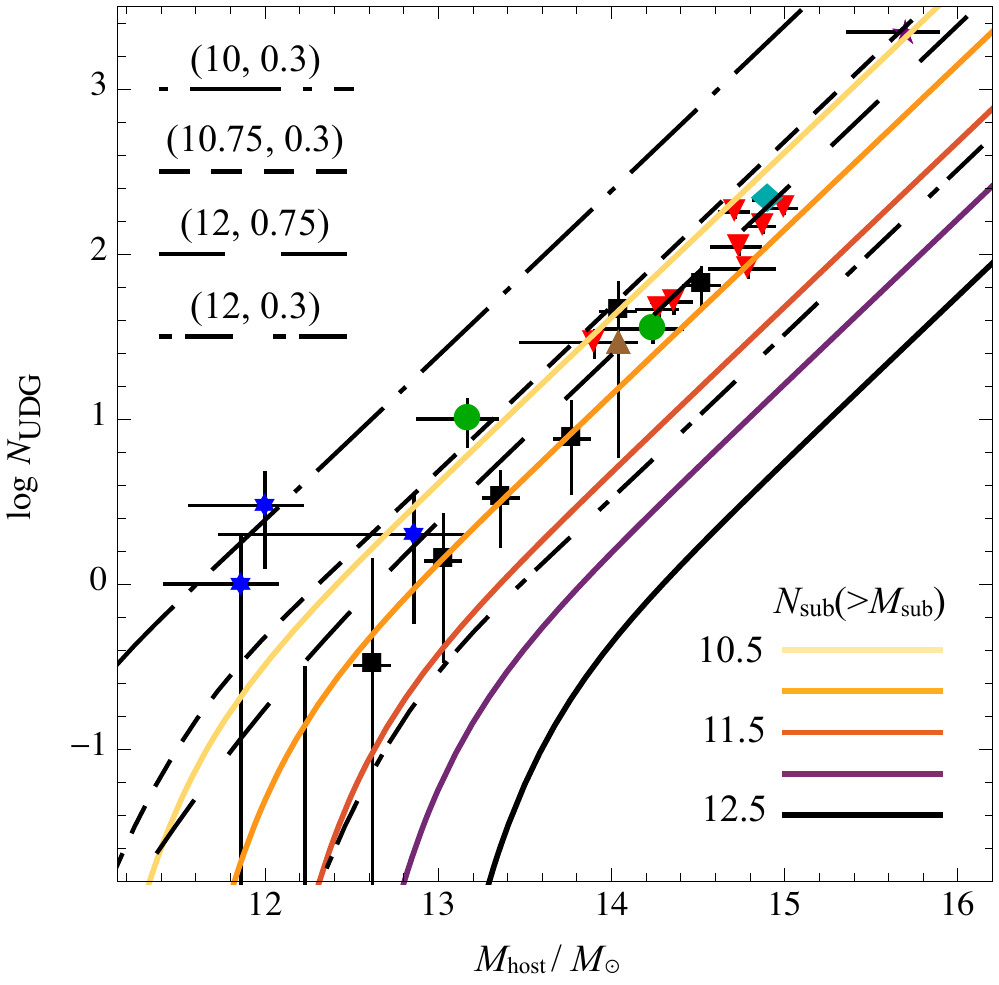}
\caption{Model UDG abundances. Coloured lines show the total number of subhaloes 
above mass $\Mh$, with values as indicated by the legend in the lower-right. Black lines with different
styles are obtained from our model for the UDG mass distribution (see text for details).}
\label{model}
\end{figure}

Before introducing a model for the fraction of UDG as a function of halo mass, in Figure~2
we compare the measured UDG abundances with the {\it total} mean number of massive subhaloes
above some threshold mass\footnote{Throughout this Letter, whenever comparing to the observed abundances, we correct model 
subhalo counts by a factor 1/0.8, to account that observations measure overdensities in cylindrical apertures. 
This correction factor assumes UDGs have an NFW spatial distribution in the cluster/group with a concentration of $c=6$
(see vdB17).}, $\langle N_{\rm sub}(>\Mh)\rangle$. Data points are the same as in Fig.~1 and 
the coloured lines display $\langle N_{\rm sub}(>\Mh)\rangle$ for the threshold masses $\log\Mh=\{10.5,11,11.5,12,12.5\}$. 
In order to roughly reproduce the observed abundances, all available subhaloes more massive than $\log\Mh\sim11$
need be occupied by UDGs. This would leave
no halo with $\log\Mh>11$ for the {non-UDG} galaxies that also populate the same clusters/groups,
showing that many UDG hosting subhaloes ought to have lower masses. In this extreme case, 
the fraction of UDGs hosted by haloes with $\Mh>10^{12}M_\odot$ is of 11.5\%. 
The same upper limit is of 39\% for the largest UDGs, with $R_e>2.5~$kpc, the abundance of which 
we estimate using the observed abundances in Fig.~2 and the size distribution measured by \citet{vdB16}.

Next, we introduce a {simple model for the fraction of subhaloes hosting UDGs, as a function of the 
present-day subhalo mass}. 
We take that the physical mechanism responsible for forming UDGs is more efficient around some 
particular mass: the fraction of {subhaloes} with mass $\Mh$ hosting UDGs has a 
gaussian shape, $\mathcal{G}(\Mh)$. Therefore, the differential UDG mass function is 
\begin{equation}
\begin{array}{ll}
{{{\rm d}\langle \NU\rangle}\over{{\rm d} \Mh}}&={{{\rm d}\langle N\rangle}\over{{\rm d} \Mh}}\ \mathcal{G}(\Mh)\\
 &={{{\rm d}\langle N\rangle}\over{{\rm d} \Mh}}\ f_{\rm max} \exp\left[-{1\over2}\left({{\log\Mh/\bar{m}}\over{\varsigma}}\right)^2\right]\ ,
 \end{array}
\label{Umf}
\end{equation}
where $\bar{m}$, $\varsigma$ and $f_{\rm max}$ are free parameters of the model. 
The parameter $\bar{m}$ is the mass at which the fraction of UDGs is largest, $f_{\rm max}$, with $f_{\rm max}\leq1$.
Notice that the value $\bar{m}$ is strictly larger than the mean UDG halo mass, and their difference 
quickly increases with the spread $\varsigma$. Due to the steepness of the subhalo mass function,
at fixed $\bar{m}$, an increase in $\varsigma$ implies higher counts of {satellite UDGs} through a larger fraction of low-mass {subhalos}.
By taking the model parameters to be constant across parent halos, by construction, 
eqn.~(\ref{Umf}) results in a linear relation between $\NU$ and $\Mc$ when $\bar{m}\ll\Mc$. 
$\NU$ may however drop below the 
linear relation when considering parent haloes with low enough mass.

Figure~2 shows the mean UDG abundances, $\langle\NU\rangle$, corresponding to our model 
mass distribution~(\ref{Umf}) for a selection of pairs $(\log\bar{m},\varsigma)$. All displayed models
adopt $f_{\rm max}=1$. By comparing with the observed abundances: 
\begin{itemize}
\item{We confirm there are more than enough low mass haloes to host the observed cluster and group UDGs. 
If $(\log \bar{m},\varsigma)=(10, 0.3)$, then a fraction $f_{\rm max}<1$ is needed. This remains 
true if $(\log \bar{m},\varsigma)=(10.75, 0.3)$, for which some deviation from linearity in the $\NU-\Mc$ relation 
can be noticed at $\log\Mc/M_\odot\lesssim12$. }
\item{A model with $(\log\bar{m},\varsigma)=(12, 0.3)$,  corresponding to a median UDG halo mass 
$\log M_{{\rm UDG},50}/M_\odot=11.8$,
cannot reproduce the observed abundances, despite $f_{\rm max}=1$. }
\item{While keeping $\log\bar{m}=12$, this can be ameliorated by increasing the value of the 
spread $\varsigma$, as shown by the model $(\log\bar{m},\varsigma)=(12, 0.75)$. This however corresponds 
to a dramatic decrease in the median UDG halo mass, with $\log M_{{\rm UDG},50}/M_\odot=10.8$.}
\end{itemize}

\section{Statistical analysis}

We now quantify the qualitative constraints above within a proper statistical framework. 
We take that, as shown by BK10, the scatter in the subhalo mass function is wider
than Poissonian, and that it approaches a fractional intrinsic scatter of $s_{\rm I}=18\%$
for large values of $\langle N\rangle$. {Here $s_{\rm I} = \sigma_{\rm I} / \langle N\rangle$,
where $\sigma_{\rm I}$ is the intrinsic scatter in $\langle N\rangle$.}  As suggested by BK10, we adopt that the probability distribution
of observing $\NU$ UDGs in a cluster of mass $\Mc$,  $P(\NU|\langle \NU\rangle(\Mc))$, is a Negative 
Binomial\footnote{{Though notice that this distribution becomes in fact sub-Poissonian for $\langle N\rangle\lesssim2$ \citep[][]{Ji17}.}} (see eqns.~(13-15) in BK10).

The observed abundances we use in this analysis are either abundances
for a single cluster or group, or mean abundances in samples of clusters or groups of similar mass. 
For the latter, we numerically construct the relevant probability distribution starting from the $P$ above 
(the parent samples are often not large enough to invoke the central limit theorem). For all of the used 
measurements we take account of the uncertainty in the group/cluster mass (as well as of the uncertainty in 
$M_{\rm WM}$, see Sect.~2.1). We take an uncertainty of 0.1~dex for the groups in vdB17 and, for the one 
group from RT17 lacking a mass uncertainty we adopt the same fractional uncertainty of the lowest-mass group in the same study. 
If, for simplicity, we still refer to the resulting probability distributions with the symbol $P$, the likelihood of the 
measured abundances $N_{{\rm UDG},i}$ is
\begin{equation}
\mathcal{L}=\prod_i P(N_{{\rm UDG},i}| \langle \NU(M_{{\rm c},i})\rangle  )
\label{lik}
\end{equation}
where the function $\langle \NU(M_{{\rm c}})\rangle$ depends on the model 
parameters $(f_{\rm max}, \bar{m}, \varsigma)$.

\begin{figure}
\centering
\includegraphics[width=.95\columnwidth]{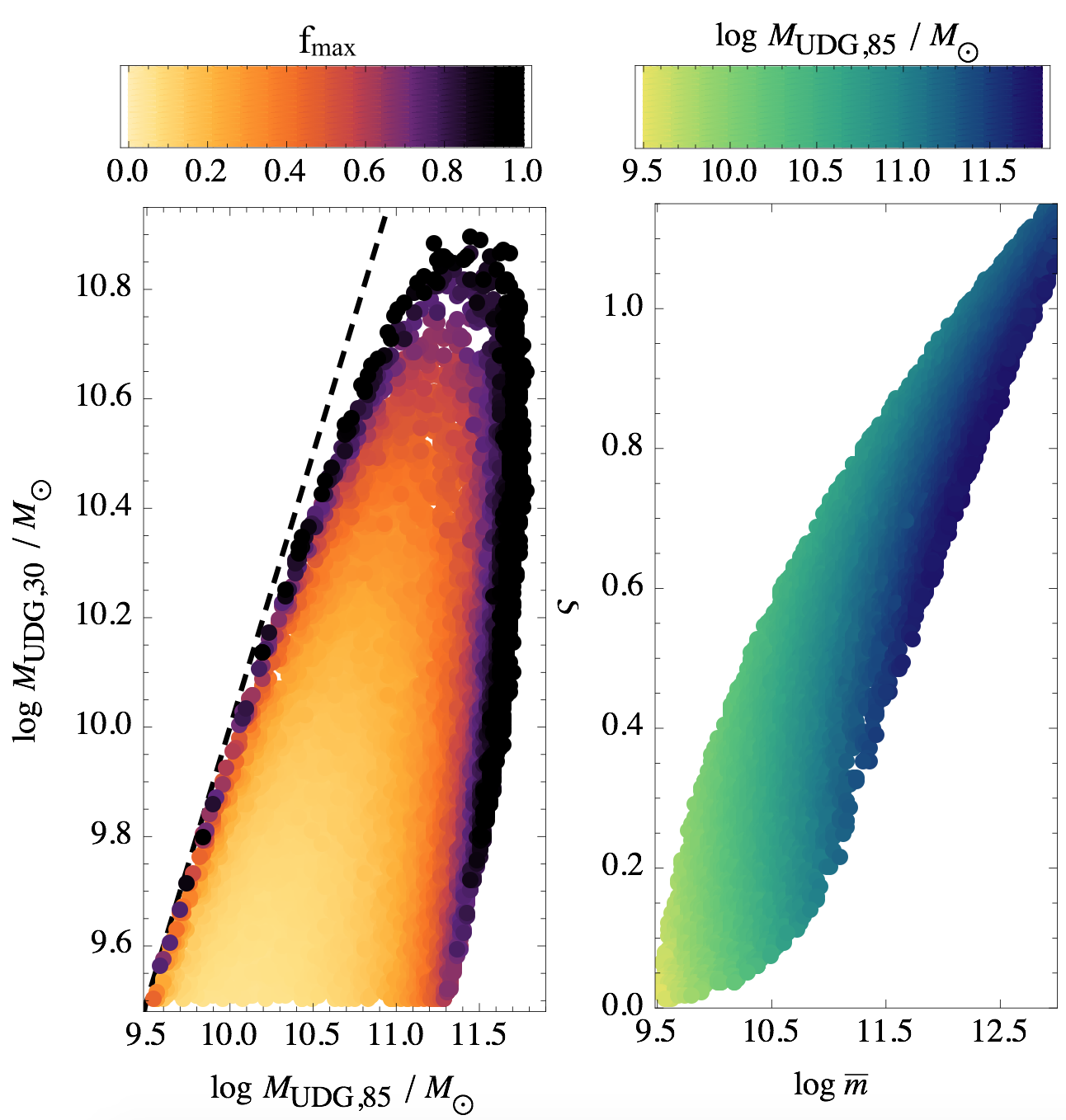}
\caption{Left panel: the distribution of accepted models in the plane of the 30\% and 85\% quantiles of the UDG virial mass distribution,
color-coded according to the maximum fraction $f_{\rm max}$; the dashed line is the one-to-one relation. Right panel: the correlation 
between the characteristic mass $\bar{m}$ and the spread $\varsigma$. The color-coding is according to the 85\% quantile of the UDG virial mass distribution. }
\label{m85}
\end{figure}

\subsection{Results}

As discussed in the previous Section, the model parameters $\bar{m}$ and $\varsigma$
are not readily interpreted. We therefore start by casting our results in terms of $M_{{\rm UDG,}30}$ and $M_{{\rm UDG,}85}$, 
respectively the 30\% and 85\% quantile of the UDG virial mass distribution. As these are a function 
of the cluster mass $\Mc$, unless otherwise specified, we take $\log\Mc=15$. In other words, we refer to the case
in which the tail at high masses of the UDG virial mass distribution is fully populated.
The left panel of Figure~3 shows the distribution of models accepted by our MCMC chains in the $(M_{{\rm UDG,}85}, M_{{\rm UDG,}30})$ 
plane. 
We only accept models that have $\log M_{{\rm UDG},30}>9.5$. 
Models that lie close to the line $M_{{\rm UDG,}85}= M_{{\rm UDG,}30}$, shown as a black dashed line,
have negligible scatter in the distribution of UDG halo masses (i.e. small values of $\varsigma$). As a consequence, 
high values of $f_{\rm max}$ are required, as indicated by the colour coding.  As $M_{{\rm UDG,}85}$ increases
from $\log M_{{\rm UDG,}85}=9.5$, even when $f_{\rm max}=1$, some minimum spread is necessary to 
reproduce the high observed abundances, and models depart from the $M_{{\rm UDG,}85}= M_{{\rm UDG,}30}$ locus. 
While $\log M_{{\rm UDG,}85}\lesssim11$, all values $0<f_{\rm max}<1$ are allowed, corresponding to
different values of $M_{{\rm UDG,}30}$, in a one-to-one relation. All of these 
models result in identical -- and very close to linear -- $\NU$-$\Mc$ relations. In this analysis these models are 
degenerate because the observed abundances do not suggest significant deviations from linearity.  
When $\log M_{{\rm UDG,}85}>11.3$, only high UDG fractions, $f_{\rm max}\gtrsim0.7$ are allowed, 
and the mass distribution is required to be broad, with $\log M_{{\rm UDG,}30}<10.9$ (or {$\log M_{{\rm UDG,}30}<11.15$ if $a=-1.86$}).
This is mirrored in the right panel of the same Figure, which shows the accepted models in the plane of
the model parameters, $(\log\bar{m}, \varsigma)$, color-coded by $M_{{\rm UDG,}85}$. The UDG abundances
alone cannot constrain these parameters, but the paucity of massive subhaloes imposes a marked correlation 
between them, corresponding to a tight upper limit on the fraction of massive UDGs.

\begin{figure}
\centering
\includegraphics[width=.95\columnwidth]{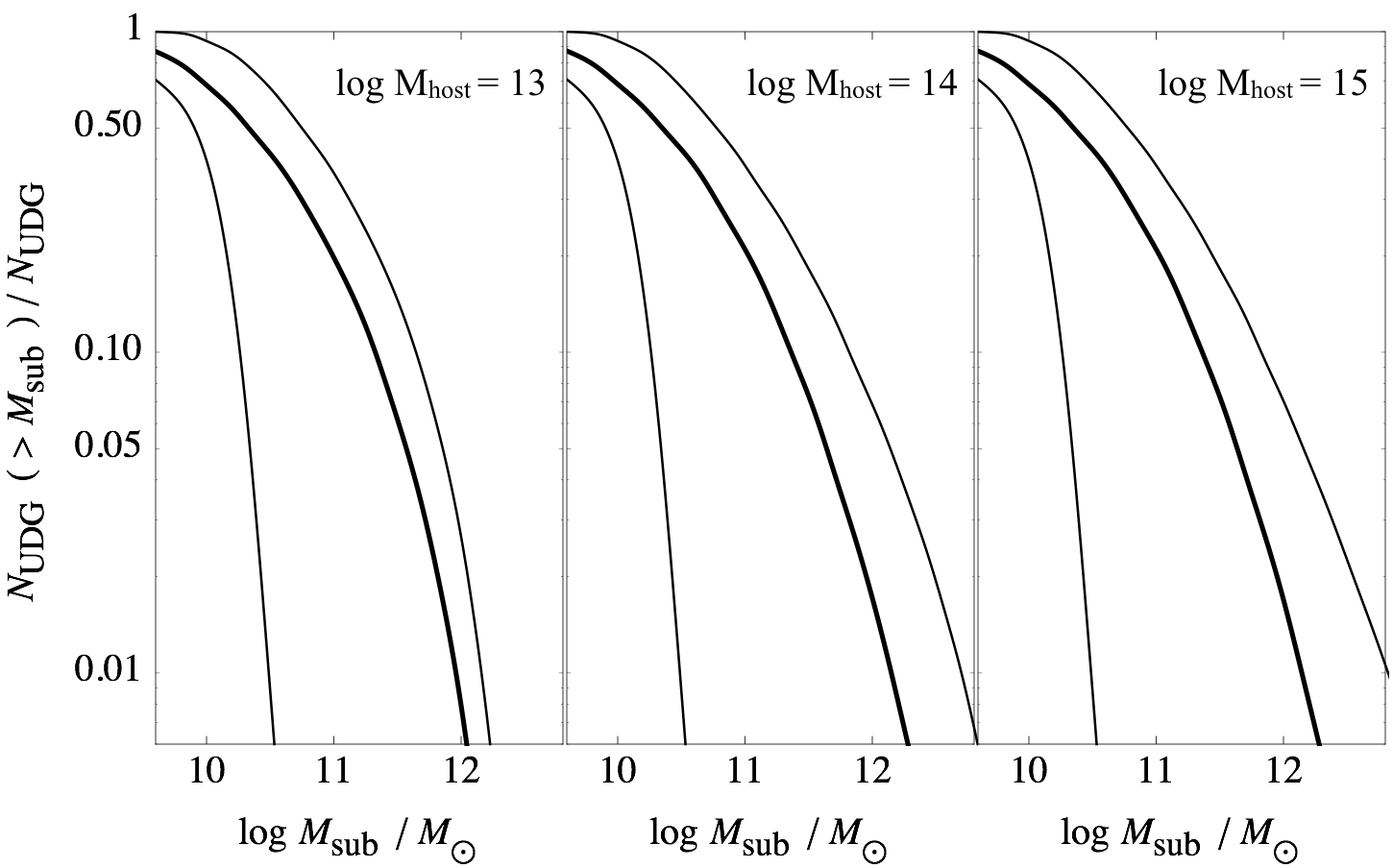}
\caption{{The cumulative distribution of the subhalo masses of UDGs in clusters with virial masses $\log\Mc/M_\odot \in\{13,14,15\}$ (respectively left, central and right panel).
The sets of lines show the 10, 50 and 90\% quantiles of the posterior distribution.
}}
\label{model}
\end{figure}

In Fig.~1, we show the resulting marginalised $\NU-\Mc$ relation. The orange shading 
identifies the 10-to-90\% confidence region for the mean UDG abundance. The data are fully consistent
with an exactly linear relation, although some deviation in low mass groups is allowed. The yellow
shaded region shows the scatter around the mean, comprising both Poisson and intrinsic scatter.
Figure~4 shows the marginalised posteriors for the cumulative mass distribution of {subhaloes} hosting UDGs (10, 50 and 90\% quantiles).
{The three panels refer to hosts of different virial mass, respectively $\log\Mc=13$, 14, and 15 in the left, central and right panels.  
In a massive cluster with $\log\Mc=15$, at 90\% probability, 50\% of all UDGs are hosted by subhaloes with $\Mh<10.8$ ({$\Mh<11.05$ if $a=-1.86$}), 
90\% by subhaloes with $\Mh<11.8$ ({$\Mh<12.1$ if $a=-1.86$}).
In a group with $\log\Mc=13$, at 90\% probability, 50\% of all UDGs are hosted by subhaloes with $\Mh<10.7$ ({$\Mh<10.8$ if $a=-1.86$}), 
90\% by subhaloes with $\Mh<11.6$ ({$\Mh<11.75$ if $a=-1.86$}).}
%
%

%
%
Fig.~5 shows the 10, 50 and 90\% quantiles for the inferred fraction of {subhaloes} 
with $\log\Mh>12$ that are occupied by UDGs, $\NU(\log\Mh>12)/N(>12)$, as a function of the 
fraction of UDGs with similarly massive subhaloes, $\NU(\log\Mh>12)/\NU$. As in Fig.~4, panels
refer to hosts with virial mass $\log\Mc/M_\odot \in\{13,14,15\}$. 
In a massive cluster with $\log\Mc=15$, if more than 5\%  ({more than 8.5\% if $a=-1.86$}) of all UDGs {have massive subhaloes, $\log\Mh>12$},
more than 50\% of subhaloes {with the same mass} are occupied by UDGs. 
{In a group with mass $\log\Mc=13$}, no accepted model has $\NU(>12)/\NU>4\%$ 
({$\NU(>12)/\NU>6\%$ if $a=-1.86$}), and if more than 2\% ({more than 2.5\% if $a=-1.86$}) of all UDGs 
are similarly massive, more than 50\% of massive haloes in groups are occupied by UDGs.

\section{Summary and Conclusions}

In this Letter, we have used the observed abundances of {satellite UDGs} in clusters and groups to 
constrain the {present-day mass distribution of their dark matter subhaloes}. If all of the most massive 
subhaloes available in the cluster host a UDG, all subhaloes with $\log\Mh/M_\odot\gtrsim11$
would be occupied by UDGs, leaving no room for the {non-UDG} galaxies in the cluster. This implies 
a sharp upper limit to the fraction of UDGs hosted by massive haloes with $\log\Mh>12$,
which is of 11.5\%. 

We introduce a model in which the efficiency of UDG formation is a function of halo mass,
{and the probability for a subhalo to host a UDG is maximum around some characteristic subhalo mass,
taken to be constant across clusters and groups. This simple assumption may more easily 
describe the scenario in which UDGs are formed in the field \citep[e.g.,][]{NA16a,AD16,Ch17} 
rather than the case in which UDGs are normal galaxies at first and expand after infall
due to satellite-specific processes such a harassment and tidal stirring.
However, we find that the currently available UDG abundances cannot constrain 
the parameters of this model, so that our specific choice has negligible impact on our results. 
Instead,} as a consequence of the limited 
number of massive subhaloes, we find that the fraction of UDGs with high virial mass 
and the spread in the UDG mass distribution are strongly correlated. For instance, if 15\% of all UDGs in 
a massive cluster have $\log\Mh>11.5$, the spread of the distribution is such that $>$30\%
has $\log\Mh<10.9$. No model in which 15\% of all UDGs in a massive cluster have $\log\Mh>11.8$
can reproduce the observed abundances. This translates in a fraction of UDGs with $\log\Mh>12$
that is $<7$\% at 90\% probability, and corresponding to a cluster in which $\sim 70$\% of all
subhaloes with $\log\Mh>12$ are occupied by UDGs. {An analysis that folds in constraints
for the fraction of satellite galaxies in clusters and groups that are UDGs vs non-UDGs is beyond the
scope of the present Letter and is left for future studies}. If we take that 50\% of all massive subhaloes
in Coma host UDGs, $<16$ out of the 332 UDGs counted by \citet{Koda15} may be massive. 
If so, the mass distribution has to be broad, with $>110$ UDGs having $\log\Mh<10.8$. 

\begin{figure}
\centering
\includegraphics[width=\columnwidth]{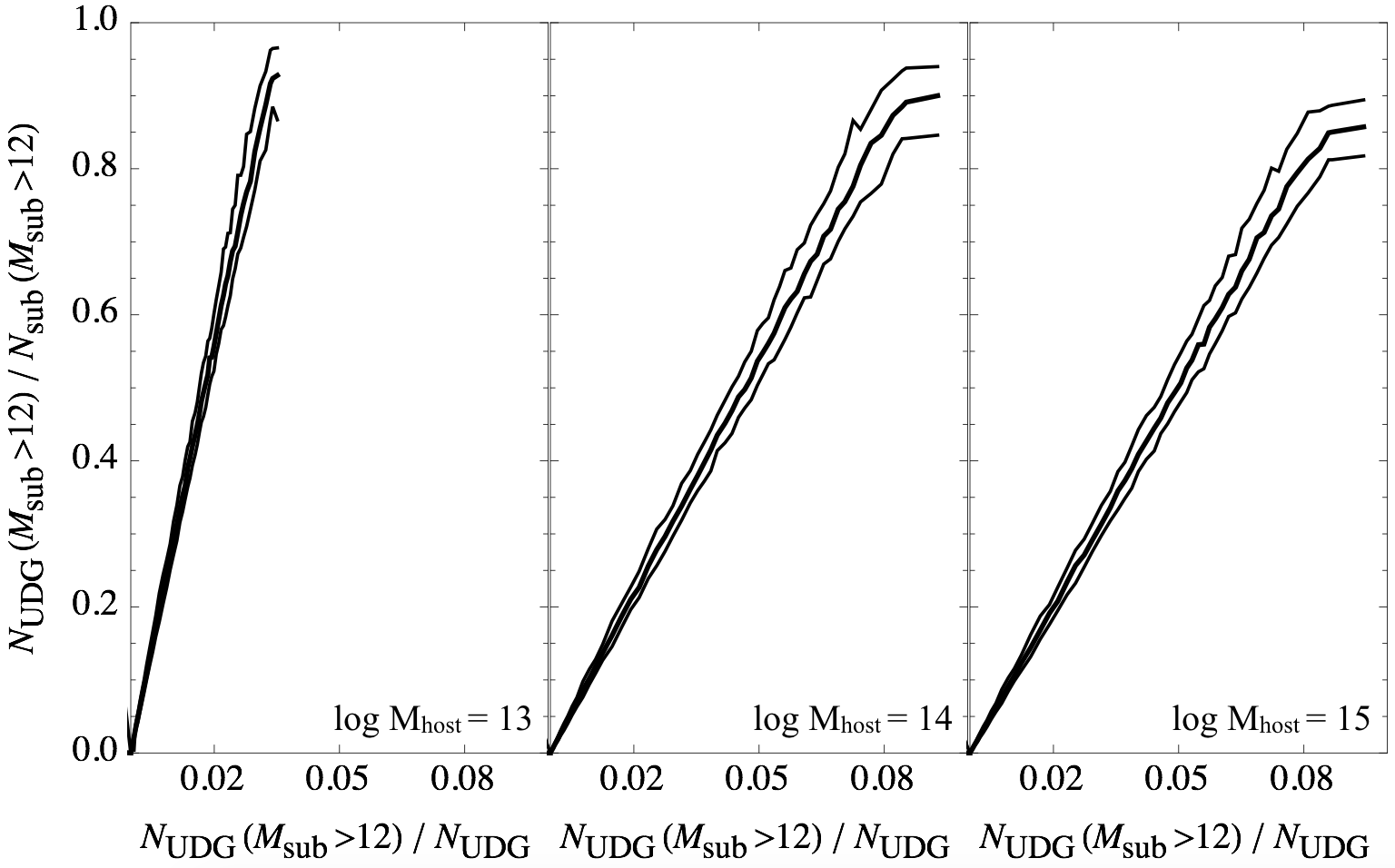}
\caption{{The mean fraction of subhaloes with present-day mass $\log\Mh>12$ that host UDGs, $\NU(\Mh>12)/N_{\rm sub}(\Mh>12)$,
against the mean fraction of satellite UDGs with similarly massive subhalos, $\NU(\Mh>12)/\NU$, as obtained from our analysis for
hosts with virial masses $\log\Mc/M_\odot \in\{13,14,15\}$ (respectively left, central and right panel).
The sets of lines show the 10, 50 and 90\% quantiles of the posterior distribution.}}
\label{model}
\end{figure}

This strongly supports a number of observational arguments suggesting that UDGs are part of a 
continuos distribution in which a majority have low mass haloes. These include: 
\begin{itemize}
\item{the seamless continuity 
of the properties of UDGs with respect to those of the numerous -- though relatively more compact -- 
LSB dwarfs \citep[e.g.,][]{Koda15,Wi17,Ve17};} 
\item{the fact that a majority of UDGs has normal GC systems for their stellar mass \citep{MB16a,MB16b,Pe16,NA16b} 
and that a minority of systems with enhanced GC abundances exist among UDGs as well as among LSB dwarfs \citep{NA16b};}
\item{the fact they do not appear to significantly deviate from the mass-metallicity relation of bright dwarf galaxies \citep{MG17,Pa17}.}
\end{itemize}

Finally, this analysis shows that it is extremely useful to better asses the properties of UDGs in low-mass groups,
as UDG abundances in this regime constrain the actual shape of the UDG virial mass distribution. We have shown that,
in proceeding towards lower mass groups, the linearity of the relation between the UDG abundance $\NU$ and 
the group mass $\Mc$ is expected to break, with mean abundances falling short of the linear relation. 
This discrepancy quantifies the weight and shape of the high mass tail of the UDG virial mass distribution.
Interestingly, the results of vdB17 appear to hint to similar deviations from linearity, with low mass groups 
($\log\Mc\sim12$) featuring a UDG in only 1 out of $\sim 10$ cases. However, as confirmed by our analysis, this 
is currently not statistically significant. Larger samples will elucidate the behaviour of the $\NU$-$\Mc$ relation 
at low group masses, allowing for better constraints on the UDG virial mass distribution and therefore 
more stringent tests for formation models.

\section*{Acknowledgements}
It is a pleasure to thank Remco van der Burg and Adriano Agnello for useful comments and the anonymous 
referee for a constructive report.


\begin{thebibliography}{99}



\bibitem[Adami et al.(2006)]{CA06} Adami, C., Scheidegger, R., Ulmer, M., et al.\ 2006, \aap, 459, 679 

\bibitem[Amorisco \& Loeb(2016)]{NA16a} Amorisco, N.~C., \& Loeb, A.\ 2016, \mnras, 459, L51 
\bibitem[Amorisco et al.(2016)]{NA16b} Amorisco, N.~C., Monachesi, A., Agnello, A., \& White, S.~D.~M.\ 2016, arXiv:1610.01595 




\bibitem[Beasley et al.(2016)]{MB16a} Beasley,  M.~A.,  Romanowsky,  A.~J.,  Pota,  V.,  et al.\ 2016,  arXiv:1602.04002 
\bibitem[Beasley \& Trujillo(2016)]{MB16b} Beasley, M.~A., \& Trujillo, I.\ 2016, arXiv:1604.08024 




\bibitem[Binggeli et al.(1985)]{BB85} Binggeli, B., Sandage, A., \& Tammann, G.~A.\ 1985, \aj, 90, 1681 


\bibitem[Boylan-Kolchin et al.(2009)]{BK09} Boylan-Kolchin, M., Springel, V., White, S.~D.~M., Jenkins, A., \& Lemson, G.\ 2009, \mnras, 398, 1150 
\bibitem[Boylan-Kolchin et al.(2010)]{BK10} Boylan-Kolchin, M., Springel, V., White, S.~D.~M., \& Jenkins, A.\ 2010, \mnras, 406, 896 











\bibitem[Chan et al.(2017)]{Ch17} Chan, T.~K., Kere{\v s}, D., Wetzel, A., et al.\ 2017, arXiv:1711.04788 

\bibitem[Conselice et al.(2002)]{CC02} Conselice, C.~J., Gallagher, J.~S., III, \& Wyse, R.~F.~G.\ 2002, \aj, 123, 2246 


\bibitem[Conselice et al.(2003)]{CC03} Conselice, C.~J., Gallagher, J.~S., III, \& Wyse, R.~F.~G.\ 2003, \aj, 125, 66 



\bibitem[Dalcanton et al.(1997)]{JD97} Dalcanton,  J.~J.,  Spergel,  D.~N.,  \& Summers,  F.~J.\ 1997,  \apj,  482,  659 
\bibitem[Davies et al.(1994)]{JD94} Davies, J., Phillipps, S., Disney, M., Boyce, P., \& Evans, R.\ 1994, \mnras, 268, 984 



\bibitem[Di Cintio et al.(2016)]{AD16} Di Cintio, A., Brook, C.~B., Dutton, A.~A., et al.\ 2016, arXiv:1608.01327 



\bibitem[Dutton et al.(2007)]{AD07} Dutton,  A.~A.,  van den Bosch,  F.~C.,  Dekel,  A.,  \& Courteau,  S.\ 2007,  \apj,  654,  27 









\bibitem[Ferguson(1989)]{HF89} Ferguson, H.~C.\ 1989, \aj, 98, 367 

\bibitem[Ferrarese et al.(2012)]{LF12} Ferrarese, L., C{\^o}t{\'e}, P., Cuillandre, J.-C., et al.\ 2012, \apjs, 200, 4 


\bibitem[Gao et al.(2004)]{Gao04} Gao, L., White, S.~D.~M., Jenkins, A., Stoehr, F., \& Springel, V.\ 2004, \mnras, 355, 819 

 


\bibitem[Giocoli et al.(2008)]{CG08} Giocoli, C., Tormen, G., \& van den Bosch, F.~C.\ 2008, \mnras, 386, 2135 



\bibitem[Greco et al.(2017)]{Gr17} Greco, J.~P., Greene, J.~E., Price-Whelan, A.~M., et al.\ 2017, arXiv:1704.06681 









\bibitem[Impey \& Bothun(1997)]{IB97} Impey, C., \& Bothun, G.\ 1997, \araa, 35, 267 



\bibitem[Janssens et al.(2017)]{Ja17} Janssens, S., Abraham, R., Brodie, J., et al.\ 2017, \apjl, 839, L17 

\bibitem[Jiang \& van den Bosch(2016)]{Ji16} Jiang, F., \& van den Bosch, F.~C.\ 2016, \mnras, 458, 2848 
\bibitem[Jiang \& van den Bosch(2017)]{Ji17} Jiang, F., \& van den Bosch, F.~C.\ 2017, \mnras, 472, 657 







\bibitem[Koda et al.(2015)]{Koda15} Koda,  J.,  Yagi,  M.,  Yamanoi,  H.,  \& Komiyama,  Y.\ 2015,  \apjl,  807,  L2 

\bibitem[Lee et al.(2017)]{Le17} Lee, M.~G., Kang, J., Lee, J.~H., \& Jang, I.~S.\ 2017, \apj, 844, 157 














\bibitem[Gu et al.(2017)]{MG17} Gu, M., Conroy, C., Law, D., et al.\ 2017, arXiv:1709.07003 









\bibitem[Mo et al.(1998)]{MMW98} Mo,  H.~J.,  Mao,  S.,  \& White,  S.~D.~M.\ 1998,  \mnras,  295,  319 




\bibitem[Mu{\~n}oz et al.(2015)]{Mun15} Mu{\~n}oz,  R.~P.,  Eigenthaler,  P.,  Puzia,  T.~H.,  et al.\ 2015,  \apjl,  813,  L15 

\bibitem[Pandya et al.(2017)]{Pa17} Pandya, V., Romanowsky, A.~J., Laine, S., et al.\ 2017, arXiv:1711.05272 




\bibitem[Peng \& Lim(2016)]{Pe16} Peng, E.~W., \& Lim, S.\ 2016, \apjl, 822, L31 

\bibitem[Penny et al.(2009)]{SP09} Penny, S.~J., Conselice, C.~J., de Rijcke, S., \& Held, E.~V.\ 2009, \mnras, 393, 1054 

\bibitem[Rom{\'a}n \& Trujillo(2017a)]{RT17a} Rom{\'a}n, J., \& Trujillo, I.\ 2017, \mnras, 468, 703 
\bibitem[Rom{\'a}n \& Trujillo(2017b)]{RT17b} Rom{\'a}n, J., \& Trujillo, I.\ 2017, \mnras, 468, 4039 








\bibitem[Sif{\'o}n et al.(2017)]{Sif17} Sif{\'o}n, C., van der Burg, R.~F.~J., Hoekstra, H., Muzzin, A., \& Herbonnet, R.\ 2017, arXiv:1704.07847 




\bibitem[Trujillo et al.(2017)]{IT17} Trujillo, I., Roman, J., Filho, M., \& S{\'a}nchez Almeida, J.\ 2017, \apj, 836, 191 




\bibitem[van den Bosch et al.(2005)]{vdB05} van den Bosch, F.~C., Tormen, G., \& Giocoli, C.\ 2005, \mnras, 359, 1029 
\bibitem[van den Bosch \& Jiang(2016)]{Bo16} van den Bosch, F.~C., \& Jiang, F.\ 2016, \mnras, 458, 2870 




\bibitem[van der Burg et al.(2016)]{vdB16} van der Burg,  R.~F.~J.,  Muzzin,  A.,  \& Hoekstra,  H.\ 2016,  arXiv:1602.00002 
\bibitem[van der Burg et al.(2017)]{vdB17} van der Burg, R.~F.~J., Hoekstra, H., Muzzin, A., et al.\ 2017, arXiv:1706.02704 



\bibitem[van Dokkum et al.(2015)]{vD15} van Dokkum,  P.~G.,  Abraham,  R.,  Merritt,  A.,  et al.\ 2015,  \apjl,  798,  L45 

\bibitem[van Dokkum et al.(2016)]{vD16} van Dokkum, P., Abraham, R., Brodie, J., et al.\ 2016, \apjl, 828, L6 

\bibitem[van Dokkum et al.(2017)]{vD17} van Dokkum, P., Abraham, R., Romanowsky, A.~J., et al.\ 2017, \apjl, 844, L11 

\bibitem[Venhola et al.(2017)]{Ve17} Venhola, A., Peletier, R., Laurikainen, E., et al.\ 2017, arXiv:1710.04616 

\bibitem[Wittmann et al.(2017)]{Wi17} Wittmann, C., Lisker, T., Ambachew Tilahun, L., et al.\ 2017, \mnras, 470, 1512 


\bibitem[Yamanoi et al.(2012)]{HY12} Yamanoi, H., Komiyama, Y., Yagi, M., et al.\ 2012, \aj, 144, 40 











\bibitem[Zaritsky(2017)]{Za16b} Zaritsky, D.\ 2017, \mnras, 464, L110 







\end{thebibliography}
\end{document}